\def\sqr#1#2{{\vcenter{\vbox{\hrule height.#2pt\hbox{\vrule
width.#2pt height#1pt \kern#1pt\vrule width.#2pt}\hrule height.#2pt}}}}

\font\cst=cmr10 scaled \magstep3

\font\csc=cmr10 scaled \magstep2
\vglue 0.5cm

\font\cst=cmr10 scaled \magstep4
\font\csc=cmr10 scaled \magstep2

\input miniltx
\input color.sty
\resetatcatcode

\centerline{ (CECS-PHY-08/21 - AEI-2008-091 - YITP-08-94)}
\vskip 0.8cm
\centerline{\cst Kerr-Schild ansatz in Einstein-Gauss-Bonnet gravity~:}
\vskip 0.5 cm
\centerline{\cst An exact vacuum solution in five dimensions}
\vskip 0.5 cm
\centerline{{\bf Andr\'es Anabal\'on}$^{1,2}$,  {\bf Nathalie Deruelle}$^{3}$, {\bf Yoshiyuki Morisawa}$^{4}$}
\bigskip
\centerline{ {\bf Julio Oliva}$^{1}$,  {\bf Misao Sasaki}$^{5}$, {\bf David Tempo}$^{1,6}$ and {\bf Ricardo
Troncoso}$^{1,7}$}

\vskip 0.5cm
\centerline{\it $^{1}$Centro de Estudios Cient\'{\i}ficos (CECS), Casilla 1469, Valdivia, Chile}
\medskip
\centerline{\it $^{2}$Max-Planck-Institut f\"{u}r Gravitationsphysik,}
\centerline{\it Albert-Einstein-Institut, Am M\"{u}hlenberg 1, DE-14476 Golm, Germany}
\medskip
\centerline{\it $^{3}$APC, UMR 7164 du CNRS, Universit\'e Paris 7}
\centerline{\it 10, rue Alice Domon et L\'eonie Duquet,
75205 Paris Cedex 13, France}
\medskip
\centerline{\it $^{4}$Faculty of Liberal Arts and Sciences, Osaka University of Economics and Law}
\centerline{\it 6-10 Gakuonji, Yao, Osaka, 581-8511, Japan}
\medskip
\centerline{\it  $^{5}$The Yukawa Institute for Theoretical Physics,}
\centerline{\it Kyoto University, Kyoto 606-8502, Japan}
\medskip
\centerline{\it $^{6}$Departamento de F\'{\i}sica, Universidad de Concepci\'{o}n,}
\centerline{\it Casilla, 160-C, Concepci\'{o}n, Chile}
\medskip
\centerline{\it $^{7}$Centro de Ingenier\'{\i}a de la Innovaci\'{o}n del CECS (CIN), Valdivia, Chile}
\footnote{}{anabalon@cecs.cl, deruelle@apc.univ-paris7.fr, morisawa@keiho-u.ac.jp,}
\footnote{}{juliooliva@cecs.cl, misao@yukawa.kyoto-.ac.jp, tempo@cecs.cl, troncoso@cecs.cl}

\vskip 1.5 true cm
\centerline{\bf Abstract}
\medskip
As is well-known, Kerr-Schild metrics linearize the Einstein tensor. We shall see here that they also simplify the Gauss-Bonnet tensor, which turns out to be only quadratic in the arbitrary Kerr-Schild function $f$ when the seed metric is maximally symmetric.
This property allows us to give a simple analytical expression for its trace,  when the seed metric  is a five dimensional maximally symmetric spacetime in spheroidal coordinates with arbitrary parameters $a$ and $b$. We also write in a (fairly) simple form the full Einstein-Gauss-Bonnet tensor (with a cosmological term) when the seed metric is flat and the oblateness parameters are equal, $a=b$.
Armed with these results we give in a compact form the solution of the trace of the Einstein-Gauss-Bonnet field equations with a cosmological term and $a\neq b$. We then examine whether this solution for the trace does solve the remaining  field equations. We find that it does not in general, unless the Gauss-Bonnet coupling is such that the field equations have a unique maximally symmetric solution.
\vskip 0.7 true cm
\centerline{*}
\vfill\eject

\vglue 0.7 true cm

{ \csc I. Introduction}
\medskip
Kerr-Schild metrics [1] are such that there exist coordinate systems $x^\mu$ in which the metric coefficients can be written as
$$g_{\mu\nu}=\overline g_{\mu\nu}+f\,h_\mu h_\nu\qquad\hbox{with}\qquad
\overline g_{\mu\nu}h^\mu h^\nu=0\quad\hbox{and}\quad h^\mu\overline D_\mu
h_\rho=0\eqno(1.1)$$
where $\overline{g}_{\mu\nu}$ are the coefficients of a seed metric in the chosen coordinates $x^\mu$ and  $f$ is an arbitrary function of the coordinates. The vector $h^\mu=\bar g^{\mu\nu}h_\nu$ is null and geodesic. Indices are moved with the seed metric $\overline{g}_{\mu\nu}$ and its inverse $\overline{g}^{\mu\nu}$, and $\overline D_\mu$ is its associated covariant derivative.

As is well-known, see e.g.  [2], the Kerr-Newman black hole solutions of Einstein's equations in four dimensions (with or without a cosmological constant) are of the Kerr-Schild type. In dimensions $D>4$, the generalization of the Kerr-(A)dS black hole solution is also of the Kerr-Schild type [3]. Note that this is not the case for the black rings [4].

In Einstein-Gauss-Bonnet (EGB) theory of gravity (see, e.g. [5] for an introduction) the spherically symmetric black hole solution found in [6] which generalizes the Schwarzschild solution is also of the Kerr-Schild type, as we shall see below.
However, as already known in the community, despite some claims to the contrary [7], and as we shall see in detail below, the Kerr-Schild ansatz which is used in [3] to obtain the $5$-dimensional Kerr (AdS) black hole solution of Einstein's equations, does not solve the EGB vacuum field equations. Some numerical results about
the existence of five-dimensional rotating black holes with angular momenta of the same magnitude have been presented in [8], and in [9] analytic results have been found up to first order in a single rotation parameter. The problem of finding a rotating black hole solution in  the EGB theory of gravity is therefore still open.

In this paper
we first study in Section 2 the properties of the Gauss-Bonnet tensor when the metric is restricted to be of the Kerr-Schild type. We find that it is quadratic in the arbitrary function $f$ (and not quartic as it could be {\sl a priori}) when the seed metric $\overline{g}_{\mu\nu}$ is maximally symmetric. This property considerably simplifies calculations. Indeed, it
 allows us to give in Section 3 a simple analytical expression for the trace of the Gauss-Bonnet tensor,  when the seed metric  is a five-dimensional maximally symmetric spacetime in spheroidal coordinates with arbitrary oblateness parameters $a$ and $b$.

   The solution of the trace of the EGB field equations with a cosmological term can hence be given  in a compact form, even when the seed metric does not solve the field equation. Furthermore,  it is shown in Section 4 that when the seed metric is restricted to be flat and when the parameters $a$ and $b$ are equal, the full EGB tensor acquires a
 (fairly) simple form.
Thus we can easily examine in this case   whether the solution for the trace obtained in Section 3 satisfies the remaining field equations.  We find that it does not, unless the Gauss-Bonnet coupling is such that there is a unique maximally symmetric solution of the field equations.
  In Section 5 we generalize this particular solution to the case when the seed metric is no longer flat and the parameters $a$ and $b$ no longer equal and we comment on a few of its properties (a detailed analysis is left to further work [18]).

\bigskip\bigskip
{ \csc II. The Einstein-Gauss-Bonnet tensor for Kerr-Schild metrics}
\medskip

The EGB tensor is [10]~:
$$E^\mu_\nu\equiv \Lambda
\delta^\mu_\nu+\kappa^{-1} G^\mu_\nu+\alpha H^\mu_\nu\,,\eqno(2.1)
$$
$G^\mu_\nu$ and $H^\mu_\nu$ being the Einstein and Gauss-Bonnet tensors respectively~:  $G^\mu_\nu\equiv R^\mu_\nu-{1\over2}R\,\delta^\mu_\nu$, and
$$H^\mu_\nu\equiv 2R^{\mu\alpha}_{\ \ \beta\gamma}R^{\beta\gamma}_{\ \
\nu\alpha} -4 R^{\mu\alpha}_{\ \ \nu\beta} R^\beta_\alpha-4R^\mu_\alpha
R^\alpha_\nu+2R R^\mu_\nu-{1\over2}\delta^\mu_\nu(R^{\alpha\beta}_{\ \
\gamma\delta}R^{\gamma\delta}_{\ \ \alpha\beta}-4R^\alpha_\beta
R^\beta_\alpha+R^2)\,.\eqno(2.2)
$$
$ R^\mu{}_{\nu\rho\sigma} \equiv 2 \partial_{[\rho} \Gamma^\mu_{\sigma]\nu} + 2 \Gamma^\mu_{\lambda[\rho} \Gamma^\lambda_{\sigma]\nu}  $,
$R_{\mu\nu}=R^\rho_{\mu\rho\nu}$ and
$R$ are the Riemann tensor, Ricci tensor and curvature scalar of the metric
$g_{\mu\nu}$. The signature is $(-++\cdots)$.  One can rewrite the cosmological constant $\Lambda$ as
$$\Lambda\equiv-{(D-1)(D-2)\over2\ell^2}\left(\kappa^{-1}-{(D-3)(D-4)\alpha\over\ell^2}\right)\,,\eqno(2.3)$$
where $D$ is the dimension of spacetime and where the two roots for $\ell^{-2}$ are the curvatures  of the maximally symmetric solutions of $E^\mu_\nu=0$.  It is worth pointing out that if the Newton constant $\kappa$ and the Gauss-Bonnet coupling $\alpha$ are related as
$$ \kappa={\ell^2\over2(D-3)(D-4)\alpha}\qquad\Longrightarrow\qquad \Lambda=-{(D-1)(D-2)(D-3)(D-4)\,\alpha\over2\ell^4}\,,\eqno(2.4)$$
then the EGB vacuum equations have a unique maximally symmetric vacuum [11], and they admit solutions with a relaxed fall-off as compared with the standard one [12]. This property enlarges the space of allowed solutions, as well as the freedom in the choice of the metric at the boundary [13].

\medskip
When the metric is of the Kerr-Schild type (1.1) the Ricci tensor $R^\mu_\nu$ is linear in $f$, as is well known [1] [3]. As for the Riemann tensors $R^\mu_{\ \nu\rho\sigma}$ and  $R^{\mu\nu}_{\ \rho\sigma}$ they are only quadratic in $f$ (see Appendix A for their explicit expressions). It also turns out that the contracted products  $R^{\mu\alpha}_{\ \ \beta\gamma}R^{\beta\gamma}_{\ \
\nu\alpha}$ and $R^{\mu\alpha}_{\ \ \nu\beta} R^\beta_\alpha$ are also quadratic in $f$ (and not respectively quartic and cubic as they could be {\sl a priori}), at least when the seed metric is maximally symmetric. Hence the result that the Gauss-Bonnet tensor $H^\mu_\nu$ is quadratic in $f$.  See Appendix A for a justification of those claims.

Let us  henceforth  restrict our attention to  anti-de Sitter
seeds with curvature ${\cal L}^{-2}$~:
$$\overline{R}_{\mu\nu\rho\sigma}=-{1\over{\cal
L}^2}(\overline{g}_{\mu\rho}\overline{g}_{\nu\sigma}-
\overline{g}_{\mu\sigma}\overline{g}_{\nu\rho})\,.\eqno(2.5)$$
(Minkowski spacetime corresponds to ${\cal L}\to\infty$ and the de Sitter case is obtained by changing the sign of ${\cal L}^2$.) Note that we do not impose the  seed metric to solve the field equations, that is, we do not impose ${\cal L}=\ell$.
\medskip
Now, at linear order in $f$,  the Einstein and Gauss-Bonnet tensors are proportional [6] (see also [14]). The remaining, quadratic, part of the Gauss-Bonnet tensor is also easily computed, so that the full EGB tensor reads
 $$\eqalign{E^\mu_\nu=&\delta^\mu_\nu(\Lambda-\Lambda_{\rm seed})+\left(\kappa^{-1}-{2(D-3)(D-4)\alpha\over{\cal
L}^2}\right)
\left[{(D-1)\over{\cal L}^2}f h^\mu
h_\nu+R^\mu_{(L)\nu}-{1\over2}\delta^\mu_\nu R_{(L)}\right]\cr
&+2\alpha\left({K\over{\cal L}^2}fh^\mu h_\nu +R^{\mu\alpha}_{(L)\beta\gamma}
R^{\beta\gamma}_{(L)\nu\alpha}-2R^{\mu\alpha}_{(L)\nu\beta}R^{\beta}_{(L)\alpha}
-2R^{\mu}_{(L)\alpha}R^\alpha_{(L)\nu}+R_{(L)}R^\mu_{(L)\nu}\right)\cr
&-{\alpha\over2}\delta^\mu_\nu\left(
R^{\alpha\beta}_{(L)\gamma\delta}R^{\gamma\delta}_{(L)\alpha\beta}-
4R^\alpha_{(L)\beta}R^\beta_{(L)\alpha}+R^2_{(L)}\right)\,,\cr}\eqno(2.6)$$
with the notation
$$\Lambda_{\rm seed}\equiv-{(D-1)(D-2)\over2{\cal L}^2}\left(\kappa^{-1}-{(D-3)(D-4)\alpha\over{\cal L}^2}\right)\,,\eqno(2.7)$$ and with the following definitions  (see Appendix A)~:
$$ R^{\mu\nu}_{(L)\rho\sigma}=\overline{g}^{\nu\alpha}
(\overline{D}_\rho\Delta^\mu_{\alpha\sigma}-\overline{D}_\sigma
\Delta^\mu_{\alpha\rho})\quad,\quad R^\mu_{(L)\nu}=\overline{g}^{\mu\sigma}\overline{D}_\rho
\Delta^\rho_{\nu\sigma}\quad,\quad R_{(L)}=\overline{D}_\rho[
h^\rho\overline{D}_\mu(fh^\mu)]\eqno(2.8)$$
with
$$\Delta^\mu_{\nu\rho}={1\over2}[\overline{D}_\nu(fh^\mu h_\rho)
+\overline{D}_\rho(fh^\mu h_\nu)-\overline{D}^\mu(fh_\nu h_\rho)]\,.\eqno(2.9)$$
As for the function $K$ it is given by
$$K\equiv 3(h^\alpha\partial_\alpha f)\overline{D}_\beta h^\beta+2(D-1)f\overline{D}_\alpha(h^\alpha\overline{D}_\beta h^\beta)+(4D-7)f\overline{D}_\alpha h^\beta(\overline{D}_\beta h^\alpha-\overline{D}^\alpha h_\beta)\,.\eqno(2.10)$$

As the Gauss-Bonnet tensor is quasilinear in the second derivatives [5], all terms quadratic in $\overline D_\lambda\partial_\rho f$ can be ignored, which simplifies the calculations. Moreover, once it has been made clear that the equations of motion are quadratic in $f$, indices in (2.6) can be moved at will using the seed metric (e.g. $R^{\mu\alpha}_{(L)\beta\gamma}
R^{\beta\gamma}_{(L)\nu\alpha}$
 can be replaced by $R^{\ \mu\ \ \beta}_{(L)\alpha\ \gamma}R_{(L)\nu\ \beta}^{\ \ \ \ \alpha\ \ \gamma}$, etc).

\bigskip\bigskip
{ \csc III. Solving the trace of the Einstein-Gauss-Bonnet equations in five dimensions}
\medskip

The general $D$-dimensional (A)dS metric in spheroidal coordinates can be found in [3]. To be specific, the 5-dimensional AdS metric in those coordinates
$x^\mu=(t,r,\theta,\phi,\psi)$ reads~:
$$\eqalign{d\overline{s}^2=-{(1+r^2/{\cal
L}^2)\Delta_\theta\over\Xi_a\Xi_b}&dt^2+
{r^2\rho^2\over(1+r^2/{\cal
L}^2)(r^2+a^2)(r^2+b^2)}dr^2+{\rho^2\over\Delta_\theta}d\theta^2+\cr
&+{r^2+a^2\over\Xi_a}\sin^2\theta\,
d\phi^2+{r^2+b^2\over\Xi_b}\cos^2\theta\,
d\psi^2\,,\cr}\eqno(3.1)$$
where $\Delta_\theta\equiv \Xi_a\cos^2\theta+\Xi_b\sin^2\theta$, with
$$\rho^2\equiv r^2+ a^2\cos^2\theta+b^2\sin^2\theta\,,\eqno(3.2)$$
 and where $\Xi_a$ and
$\Xi_b$ are related to the parameters $a$ and $b$ by
$\Xi_a\equiv 1-a^2/{\cal L}^2$, $\Xi_b\equiv 1-b^2/{\cal L}^2$. As for
the null and geodesic vector
$h_\mu$ it is given by
$$h_\mu
dx^\mu={\Delta_\theta\over\Xi_a\Xi_b}\,dt+{r^2\rho^2\over(1+r^2/{\cal
L}^2)(r^2+a^2)(r^2+b^2)}\,dr+{a\sin^2\theta\over\Xi_a}d\phi+{b\cos^2\theta\over\Xi_b}d\psi\,.\eqno(3.3)$$

Using the properties listed in the preceding section, and imposing $f$ to depend on $r$ and $\theta$ only,  the trace of the EGB tensor in these coordinates is easily calculated  and turns out to have the following remarkably simple form~:
$$E=5(\Lambda-\Lambda_{\rm seed})-{(rQ_t)''\over 2r \rho^2}\quad\hbox{with}\quad Q_t=\left(\kappa^{-1}-{4\alpha\over{\cal L}^2}\right)Q_l+\alpha\, {Q_q}\quad\hbox{and}\quad\left\{\eqalign{Q_l&=3\rho^2f\cr Q_q&=2(4r^2-\rho^2){f^2\over\rho^2}\,,\cr}\right.\eqno(3.4)$$
prime denoting derivation with respect to $r$.

It is now a simple matter to solve the vacuum EGB equation for the trace, $E=0$~:
$$Q_t=6\,m(\theta)+{d(\theta)\over r}+{(\Lambda-\Lambda_{\rm seed})\over6}\,r^2\,(10\rho^2-7r^2)\,,\eqno(3.5)$$
where $m(\theta)$ and $d(\theta)$ are arbitrary functions of $\theta$.
The solutions for $f$ are therefore the roots of the second degree equation~:
$$2\alpha(4r^2-\rho^2){f^2\over\rho^2}+3\left(\kappa^{-1}-{4\alpha\over{\cal L}^2}\right)\rho^2f=6\,m(\theta)+{d(\theta)\over r}+{(\Lambda-\Lambda_{\rm seed})\over6}\,r^2\,(10\rho^2-7r^2)\,.\eqno(3.6)$$

We have now to check whether this solution for the trace satisfies the  remaining field equations. As we shall first see explicitly in the next section, in the particular case when ${\cal L}\to\infty$ and $a=b$, the answer is ``no" unless the Gauss-Bonnet coupling is fixed as in Eq. (2.4).
\bigskip\bigskip
{ \csc IV. Flat seed metric and equal oblateness parameters}
\medskip
{\bf The EGB tensor in terms of $f$:} In contrast to the trace, it is a more painstaking task to express the full EGB tensor $E^\mu_\nu$ in a simple manner. Thus, as a warming up exercise, we shall restrict  our attention in this section to the simple case ${\cal L}\to\infty$ and $a=b$  (the case $a=-b$ is trivially obtained by a parity transformation).
The (flat) seed metric then simply reads
$$d\overline{s}^2=-dt^2+{r^2\over r^2+a^2}dr^2+(r^2+a^2)(d\theta^2+\sin^2\theta d\phi^2+\cos^2\theta d\psi^2)\,,\eqno(4.1)$$
and the null and geodesic vector is
$$h_\mu=\left(1,\ \ {r^2\over r^2+a^2}\ \,\ 0\ ,\ a\sin^2\theta\ ,\  a\cos^2\theta\right)\,.\eqno(4.2)$$

The trace of the EGB tensor reduces to, see (3.4) and (3.2)~:
$$E=5\Lambda-{(rQ_t)^{''}\over2r(r^2+a^2)}\quad\hbox{with}\quad Q_t=\kappa^{-1}Q_l+\alpha Q_q\quad\hbox{where}\quad\left\{\eqalign{Q_l&=3(r^2+a^2)f\cr Q_q&={2(3r^2-a^2)\over r^2+a^2} f^2\,,\cr}\right.\eqno(4.3)$$
and the solution of $E=0$ is~:
$$Q_t=6\,m+{d\over r}+{\Lambda\over6}r^2(3r^2+10a^2)\,.\eqno(4.4)$$
Since $Q_t$ is known in terms of $f$ (see (4.3)), $f$  hence solves the trace equation if it is a root of
$$2\alpha{(3r^2-a^2)\over(r^2+a^2)}f^2+3\kappa^{-1}(r^2+a^2)f=6\,m+{d\over r}+{\Lambda\over6}r^2(3r^2+10a^2)\,.\eqno(4.5)$$

Careful examination then shows that all components of $E^\mu_\nu$ can be expressed in terms of $E^r_r$ and $E^\phi_\psi$ as~:
$$ \eqalign{ E^t_t&=-{a^2\over3(r^2+a^2)}\left({a^2+r^2\over r}{E^r_r}'+{2E^\phi_\psi\over\cos^2\theta}\right)+E^r_r\cr
E^t_\phi&=-{a\sin^2\theta\over3}\left({a^2+r^2\over r}{E^r_r}'+{2E^\phi_\psi\over\cos^2\theta}\right)\cr
E^t_\psi&=-{a\cos^2\theta\over3}\left({a^2+r^2\over r}{E^r_r}'+{2E^\phi_\psi\over\cos^2\theta}\right)\cr
E^\theta_\theta&={1\over3}\left({a^2+r^2\over r}{E^r_r}'-{E^\phi_\psi\over\cos^2\theta}\right)+E^r_r\cr
E^\phi_\phi&={1\over3}\left({a^2+r^2\over r}{E^r_r}'+(2-3\cos^2\theta){E^\phi_\psi\over\cos^2\theta}\right)+E^r_r\cr
E^\psi_\psi&={1\over3}\left({a^2+r^2\over r}{E^r_r}'-(1-3\cos^2\theta){E^\phi_\psi\over\cos^2\theta}\right)+E^r_r\,,\cr}   \eqno(4.6)$$
all other components being either zero or obtained by raising / lowering indices with the seed metric.
As for $E^r_r$ and $E^\phi_\psi$ they are expressed in terms of the function $f$ or, rather, $Q_t$ and $Q_q$, see (4.3), as
 $$E^r_r=\Lambda+{1\over 6r(r^2+a^2)^2}\left[-(3r^2+a^2)Q^{\prime}_t+8\alpha\, a^4\left({Q_q\over3r^2-a^2}\right)^{\prime}\right]\,,\eqno(4.7)$$
and (an admittedly ugly expression)
$$\eqalign{{E^\phi_\psi\over\cos^2\theta}=&{a^2[(a^2+5r^2)Q^{\prime}_t-r(r^2+a^2)Q''_t]\over6r^3(r^2+a^2)^2}\cr
&+{4\alpha a^2(27r^4+42r^2a^2+31a^4)\over(3r^2-a^2)^3(r^2+a^2)^2}Q_q\cr
&-{4\alpha a^2(18r^6+27r^4a^2+16r^2a^4-a^6)\over3r^3(3r^2-a^2)^2(r^2+a^2)^2}Q_q^{\prime}\cr
&+{2\alpha a^2(3r^2+2a^2)\over3r^2(3r^2-a^2)(r^2+a^2)}Q_q''\,.\cr}\eqno(4.8)$$

\bigskip
{\bf Recovering the Boulware-Deser metric ($a=b=0$).}
In the case of vanishing oblateness parameters, that is, when the Minkowski seed metric (4.1) is written is standard spherical coordinates, the non-zero components of the $5$-D EGB tensor (4.6-4.8) simplify into
$$E^t_t=E^r_r=-{Q_t'\over2r^3}+\Lambda\quad,\quad E^\theta_\theta=E^\phi_\phi=E^\psi_\psi={[r^3E^r_r]'\over3r^2}\,.\eqno(4.9)$$
Thus, the field equations $E^\mu_\nu=0$ are solved by~:
$$Q_t=6m+{\Lambda\over2}\,r^4\,.\eqno(4.10)$$
It then follows from Eq. (4.3) that the function $f$ is given by
$$f(r)={r^2\over4\kappa\alpha}\left(-1\pm\sqrt{1+{8  \kappa^{2}\alpha \over3 r^4}\left(6m+{\Lambda r^4\over2}\right)}\right)\,,\eqno(4.11)$$
which is nothing but the EGB solution first found in [6], written here in Kerr-Schild form. (The equivalence of the metrics follows from the generalized Birkhoff theorem [14], and can be explicitly seen from the coordinate transformation, $t=T+\int{f(r)\over 1-f(r)}dr$, which brings it into the Schwarzschild gauge.)  The constant of integration
$m$ is interpreted as the total mass, see [6] [15-17].
\bigskip
{\bf Switching on the oblateness parameters $(a=b\neq 0)$. } In this case the solutions of the trace equation are the roots of (4.5), that is~:
$$f(r)={3(r^2+a^2)^2\over4\kappa\alpha(3r^2-a^2)}\left(-1\pm\sqrt{1+{8\kappa^2\alpha(3r^2-a^2)\over9(r^2+a^2)^3}\left(6m+{d\over r}+{\Lambda\over6}r^2(3r^2+10a^2)\right)}\right)\,.\eqno(4.12)
$$
The Kerr-Schild metric so obtained seems to describe a massive rotating spacetime. However
one can  show that, with $f(r)$ given by (4.12), the equations of motion $E^\mu_\nu=0$ with $E^\mu_\nu$ given by (4.6-4.8) cannot be satisfied, unless the Gauss-Bonnet coupling is fixed as in Eq. (2.4).

 This can be seen from the behavior of $E^r_r$ when $r\to\infty$, see Eq. (4.7)~:
$$E^r_r={2a^4\over 9\kappa^2\alpha r^4}\sqrt{3+4\kappa^2\alpha\Lambda}\left(\sqrt{3+4\kappa^2\alpha\Lambda}\mp\sqrt{3}\right)+{d\over2r^5}+\cdots \,.\eqno(4.13)$$
The leading term of Eq. (4.13) vanishes either for $\Lambda=0$ taking the upper branch of (4.12), or for $\alpha=-{3\over 4\kappa^2\Lambda}$, and the subleading term vanishes for $d=0$.

In the first case ($\Lambda=0$,  $d=0$), the asymptotic behavior of $E^r_r$ becomes
$$E^r_r=-{64 a^4\alpha \kappa^2m^2\over r^{12}}+\cdots \,,\eqno(4.14)$$
which means that $m$ also must vanish. Hence $f=0$ and the solution reduces to the flat seed metric.

In the case now  when $\alpha=-{3\over 4\kappa^2\Lambda}$ (and $d=0$), the equation (4.7) $E_r^r=0$ is fulfilled provided $m={7a^4\Lambda\over36}$, and the solution for $f$ in (4.12)
 has the remarkably simple form
$$f(r)=-{r^2+a^2\over\ell^2}\,,\eqno(4.15)$$
where $\ell^{-2}=(4\alpha\kappa)^{-1}$ stands for the (A)dS curvature of the unique maximally symmetric solution of the field equations. It is an exercise to check that $E^\phi_\psi$ given in (4.8) also vanishes, so that all the components of the EGB equations, $E^\mu_\nu=0$ with $E^\mu_\nu$ given by
(4.6) are indeed satisfied.

Thus  the Kerr-Schild solution with flat seed metric (4.1) and function $f(r)$ given by (4.15) solves the EGB field equations when $\alpha=-{3\over 4\kappa^2\Lambda}$ (and hence $\ell^{2}=4\alpha\kappa)$.

It is worth pointing out that this solution is static, since there is a coordinate transformation that brings it into the Schwarzschild gauge, but where the three-sphere is replaced by a squashed three-sphere, where $a$ parameterizes the squashing [18]. This is not at odds with the Birkhoff  theorem [14] since the freedom in the choice of the metric at the boundary is enlarged when the Gauss-Bonnet coupling is fixed as in Eq. (2.4) [13].

Let us now generalize the previous analysis to the case when the seed metric is no longer flat and $b$ is no longer equal to $a$.

\bigskip\bigskip
{ \csc V. An exact vacuum solution}
\medskip
When the seed metric is no longer flat and $b$ is no longer equal to $a$, that is when the anti-de Sitter seed metric is given by (3.1), we can no longer give simple expressions for the components
of the EGB tensor as we did in the previous section. However we know the general solution for the trace, see (3.6)~:
$$f(r,\theta)=A(-1\pm\sqrt{1+B})\quad\hbox{with}\quad\left\{\eqalign{A&={3(\kappa^{-1}-4\alpha/{\cal L}^2)\rho^4\over4\alpha(4r^2-\rho^2)}\quad,\quad B={8\alpha(4r^2-\rho^2)\over9\rho^6(\kappa^{-1}-4\alpha/{\cal L}^2)^2}Q_t\cr
Q_t&=\left(6\,m(\theta)+{d(\theta)\over r}+{(\Lambda-\Lambda_{\rm seed})\over6}\,r^2\,(10\rho^2-7r^2)\right)\cr}\right.\eqno(5.1)$$
where we recall that $\rho^2=r^2+a^2\cos^2\theta+b^2\sin^2\theta$.
\medskip
Let us first extend the ``no-go" result of the preceding section~:
if $a$ and/or $b$ are not zero and if one chooses  $\Lambda=\Lambda_{\rm seed}$, (i.e. the seed metric solves the vacuum field equations) then, due to the properties of the Gauss-Bonnet tensor listed in Appendix A, it is possible to show that
the solution (5.1) for the trace does not solve the remaining field equations, even in the particular case when $\kappa^{-1}={4\alpha/\ell^2}$ (unless $f=0$). Therefore the Kerr-Schild ansatz which was used in [3] to obtain the $5$-dimensional Kerr (AdS) black hole solution of Einstein's equations, does not yield a rotating solution of the EGB vacuum field equations.
\medskip
Let us now generalize the solution obtained in the preceding section. For $\Lambda\neq\Lambda_{\rm seed}$ and the Gauss-Bonnet coupling  fixed as in Eq. (2.4), the function $f$ is simply given by
$$f(r,\theta)=-\left({1\over\ell^2}-{1\over{\cal L}^2}\right)\rho^2\,.\eqno(5.2)$$
Thus, in five dimensions, the line element
$$ds^2=d\overline{s}^2-\left({1\over\ell^2}-{1\over{\cal L}^2}\right)\rho^2(h_\mu dx^\mu)^2\,,\eqno(5.3)$$
where $d\overline{s}^2$ is the metric of a seed (anti)-de Sitter spacetime of curvature ${\cal L}^{-2}$ in spheroidal coordinates and $h_\mu$ is a null and geodesic vector, see Eqs. (3.1) and (3.3), solves the EGB vacuum equations, if the Gauss-Bonnet coupling is fixed in such a way that the field equations have a unique maximally symmetric solution of curvature  $\ell^{-2}$, see (2.4).

This spacetime is asymptotically locally anti-de Sitter since
$$R^{\alpha\beta}_{\ \ \gamma\delta}\to-{1\over\ell^2}(\delta^\alpha_\gamma\delta^\beta_\delta-\delta^\alpha_\delta\delta^\beta_\gamma)\quad\hbox{when}\quad r\to\infty\,.\eqno(5.4)$$
Note that it does not approach the seed metric since ${\cal L}$ must be different from $\ell$.
The solution is parameterized by three constants~: ${\cal L}^{-2}$, that is, the curvature of the seed (anti)-de Sitter spacetime, and the parameters $a$ and $b$ defining  the spheroidal coordinates. Preliminary results show that this solution describes a rotating spacetime [18] if $a\neq b$, but a detailed analysis is left to further work.

\bigskip\bigskip
{ \csc VI. Conclusions}
\medskip
We considered Kerr-Schild type metrics on maximally symmetric seed spacetimes and showed that  the EGB tensor is only quadratic in the Kerr-Schild function $f$, see Eq (2.6). Specializing to a $5$-dimensional seed metric in spheroidal coordinates we then found a remarkably simple expression for the trace of the Einstein Gauss-Bonnet tensor, see Eq (3.4). Specializing further to a flat seed metric and equal spheroidal parameters we wrote explicitly all the components of the EGB tensor, see Eq (4.6-4.8). Thanks to those results we were able to show in a transparent manner that the  Kerr-Schild ansatz used in [3] to obtain the generalized 5-dimensional Kerr solution in Einstein theory
does not yield a solution of the EGB vacuum equations, when, as in [3], the (anti)-de Sitter seed metric is chosen so as to solve the field equations.
Turning then to Kerr-Schild ansatze whose (anti)-de Sitter seed metric does not solve the field equations, see Eq (3.1-3.3), we found a new solution given by (5.3) provided the Gauss-Bonnet coupling is fixed as in Eq. (2.4).

\vskip 0.5cm
{\bf Acknowledgements.}
We thank Alex Giacomini and Hideki Maeda for enlightening discussions. N.D. thanks Gary Gibbons for useful correspondence and the Yukawa Institute in Kyoto and CECS in Valdivia  for their hospitality. A.A. is an Alexander von Humboldt fellow. Y.M. is supported by the OUEL research fund. D.T. thanks CONICYT for financial support. This work was partially funded by FONDECYT grants 1051056, 1061291, 1071125, 1085322, 3085043, 3080024, and by the international collaboration project ACI 01 (PBCT-Chile); The Centro de Estudios Cient\'{\i}ficos (CECS) is funded by the Chilean Government through the Millennium Science Initiative and the Centers of Excellence Base Financing Program of CONICYT. CECS is also supported by a group of private companies which at present includes Antofagasta Minerals, Arauco, Empresas CMPC, Indura, Naviera Ultragas and Telef\'{o}nica del Sur. CIN is funded by CONICYT and the Gobierno Regional de Los R\'{\i}os.
This work was also supported in part by JSPS Grant-in-Aid for Scientific
Research (B) No.~17340075, and (A) No.~18204024, by JSPS
Grant-in-Aid for Creative Scientific Research No.~19GS0219,
and by Monbukagaku-sho Grant-in-Aid for the global COE program,
``The Next Generation of Physics, Spun from Universality and Emergence".

\vskip 0.5cm

\noindent {\csc References}
\medskip
 \item{[1]}
 R.P. Kerr and A. Schild,  Proc. Symp. Appl. Math. 17, 199 (1965)

G.C. Debney, R.P. Kerr and A. Schild, J. Math. Phys. 10 (1969), 245 

\item{[2]}  ``Exact solutions of Einstein's field equations", 2nd edition, H. Stephani et al., Cambridge University Press, 2003

\item{[3]} R.C. Myers and M.J. Perry,  Ann. Phys.
172, 304 (1986) 

S.W. Hawking, C.J. Hunter, M. Taylor, Phys. Rev. {\bf D59} 064005 (1999). 

 G.W. Gibbons, H. Lu, D.N. Page and C.N. Pope, J.\ Geom.\ Phys.\  {\bf 53}, 49 (2005) 

\item{[4]}
V.~Pravda, A.~Pravdova and M.~Ortaggio, Class.\ Quant.\ Grav.\  {\bf 24}, 4407 (2007), 
 
R. Emparan and H.S. Reall, Living Rev.\ Rel.\  {\bf 11}, 6 (2008), 
 
M.~Ortaggio, V.~Pravda and A.~Pravdova, Class.\ Quant.\ Grav.\  {\bf 26}, 025008 (2009). 

\item{[5]} N. Deruelle and J. Madore, ``On the quasi-linearity of the Einstein- 'Gauss-Bonnet' gravity field equations,''  arXiv:gr-qc/0305004.

C. Garraffo and G. Giribet, Mod.\ Phys.\ Lett.\  A {\bf 23}, 1801 (2008) 

\item{[6]} D.G. Boulware and S. Deser, Phys. Rev. Lett. 55 (1985) 2656 

J. Wheeler,  Nucl. Phys. B268 (1986) 737 
; Nucl. Phys. B273 (1986) 732  

D. Wiltshire,  Phys. Lett. B169 (1986) 36.   

For this theory, black holes with horizons of constant curvature have also been found in Refs.

R.~Aros, R.~Troncoso and J.~Zanelli, Phys.\ Rev.\  D {\bf 63}, 084015 (2001),  

R.~G.~Cai, Phys.\ Rev.\  D {\bf 65}, 084014 (2002),  

R.~G.~Cai and Q.~Guo, Phys.\ Rev.\  D {\bf 69}, 104025 (2004).  

\item{[7]}
S. Alexeyev,
N. Popov, M. Startseva, A. Barrau and J. Grain, J.\ Exp.\ Theor.\ Phys.\  {\bf 106}, 709 (2008)  

\item{[8]} Y. Brihaye, E. Radu, Phys.\ Lett.\  B {\bf 661}, 167 (2008)

\item{[9]} H. Kim, R.G. Cai, Phys.\ Rev.\  D {\bf 77}, 024045 (2008)

\item{[10]} C. Lanczos, Z. Phys., 73 (1932) 147

C. Lanczos, Annals Math.\  {\bf 39}, 842 (1938).

D. Lovelock, J.\ Math.\ Phys.\  {\bf 12}, 498 (1971).

\item{[11]}   J.~Crisostomo, R.~Troncoso and J.~Zanelli,
  Phys.\ Rev.\  D {\bf 62}, 084013 (2000).

\item{[12]} M. Henneaux, C. Teitelboim, Commun.\ Math.\ Phys.\  {\bf 98}, 391 (1985).

 \item{} M. Henneaux, ``Asymptotically Anti-De Sitter Universes In D = 3, 4
And Higher Dimensions", Proceedings of the Fourth Marcel Grossmann
Meeting on General Relativity, Rome 1985. R. Ruffini (Ed.), Elsevier
Science Publishers B.V., pp. 959-966

\item{[13]} G. Dotti, J. Oliva, R. Troncoso, Phys.\ Rev.\  D {\bf 76}, 064038 (2007).
; G. Dotti, J. Oliva, R. Troncoso, ``Vacuum solutions with nontrivial boundaries for the Einstein-Gauss-Bonnet theory",  arXiv:0809.4378 [hep-th]. CECS-PHY-08-12, To appear in the proceedings of 7th Alexander Friedmann International Seminar on Gravitation and Cosmology, Joao Pessoa, Brazil.

\item{[14]} C. Charmousis and J.F. Dufaux, Class.\ Quant.\ Grav.\  {\bf 19}, 4671 (2002)

R. Zegers, J.\ Math.\ Phys.\  {\bf 46}, 072502 (2005).

S. Deser and J. Franklin, Class.\ Quant.\ Grav.\  {\bf 22}, L103 (2005).

\item{[15]} S. Deser and B. Tekin,   Phys.\ Rev.\  D {\bf 67}, 084009 (2003).

 A. Padilla, Class.\ Quant.\ Grav.\  {\bf 20}, 3129 (2003).

 N. Deruelle, J. Katz and S. Ogushi, Class.\ Quant.\ Grav.\  {\bf 21}, 1971 (2004).

\item{[16]} N. Deruelle and Y. Morisawa, Class.\ Quant.\ Grav.\  {\bf 22}, 933 (2005).

N. Deruelle, ``Mass and angular momenta of Kerr-anti-de Sitter spacetimes,''
  arXiv:gr-qc/0502072.

 \item{[17]}  P. Mora, R. Olea, R. Troncoso and J. Zanelli, JHEP {\bf 0406}, 036 (2004).

 \item{[18]} A. Anabal\'on, N. Deruelle, J. Oliva, D. Tempo, R. Troncoso, ``Exact rotating solution in vacuum for the Einstein-Gauss-Bonnet theory in five dimensions", CECS-PHY-08/19

  \vskip 0.5cm
{\bf Appendix A. Some properties of Kerr-Schild metrics}
\medskip

Kerr-Shild metrics read
$$g_{\mu\nu}=\overline g_{\mu\nu}+f\,h_\mu h_\nu\qquad\hbox{with}\qquad \overline g^{\mu\nu}h_\mu h_\nu=0\qquad\hbox{and}\qquad h^\mu\overline D_\mu
h^\rho=0\,.$$
The function $f$ is arbitrary and $h^\nu\equiv\overline{g}^{\mu\nu}h_\mu$. An overlined
quantity is built with the seed metric $\overline g_{\mu\nu}$.
\medskip

The Christoffel symbols are given by
$$\Gamma^\mu_{\nu\rho}-\overline{\Gamma}^\mu_{\nu\rho}=
\Delta^\mu_{\nu\rho}+\delta^\mu_{\nu\rho}$$
with
$$\Delta^\mu_{\nu\rho}={1\over2}[\overline{D}_\nu(fh^\mu h_\rho)
+\overline{D}_\rho(fh^\mu h_\nu)-\overline{D}^\mu(fh_\nu h_\rho)]\eqno({\rm A}.1)$$
and
$$\delta^\mu_{\nu\rho}={1\over2}h^\mu h_\nu
h_\rho(fh^\lambda\partial_\lambda f)\,.$$
$\Delta^\mu_{\nu\rho}$ has the following properties
$$\Delta^\rho_{\nu\rho}=0\ \ ,\ \
h^\rho\Delta^\mu_{\nu\rho}={1\over2}h^\mu h_\nu (h^\rho\partial_\rho f)
\   ,\  \ h_\mu\Delta^\mu_{\nu\rho}=-{1\over2}h_\nu h_\rho
(h^\mu\partial_\mu f)\ \ ,\ \
\Delta^\alpha_{\nu\beta}\Delta^\beta_{\nu\alpha}\propto
h_\mu h_\nu\,.$$

\medskip
It follows that the Riemann tensor, which {\sl a priori} could
be quartic in $f$, is in fact only quadratic and reads
$$R^\mu_{\ \nu\rho\sigma}= \overline{R}^\mu_{\ \nu\rho\sigma}+
R^\mu_{(lin)\, \nu\rho\sigma}+R^\mu_{(quad)\,\nu\rho\sigma}$$
with
$$R^\mu_{(lin)\,
\nu\rho\sigma}=\overline{D}_\rho\Delta^\mu_{\nu\sigma}
-\overline{D}_\sigma\Delta^\mu_{\nu\rho}$$ and
$$R^\mu_{(quad)\,\nu\rho\sigma}=\overline{D}_\rho\delta^\mu_{\nu\sigma}
-\overline{D}_\sigma\delta^\mu_{\nu\rho}
+\Delta^\mu_{\rho\lambda}\Delta^\lambda_{\nu\sigma}-
\Delta^\mu_{\sigma\lambda}\Delta^\lambda_{\nu\rho}\,.$$
$R^\mu_{(quad)\,\nu\rho\sigma}$ has the following properties
$$h^\nu R^\mu_{(quad)\,\nu\rho\sigma}=0\quad,\quad
h_\mu R^\mu_{(quad)\,\nu\rho\sigma}=0\quad,\quad
h^\sigma R^\mu_{(quad)\,\nu\rho\sigma}=-{1\over2}h^\mu h_\nu h_\rho (f h^\alpha
h^\beta \overline{D}_{\alpha\beta} f)\,.$$
$R^\mu_{(lin)\,\nu\rho\sigma}$ has the following properties
$$h^\nu h^\sigma R^\mu_{(lin)\,\nu\rho\sigma}=-{1\over2}h^\mu h_\rho (h^\alpha
h^\beta\overline{D}_{\alpha\beta} f)\quad,\quad
h_\mu h^\sigma R^\mu_{(lin)\,\nu\rho\sigma}={1\over2}h_\nu h_\rho (h^\alpha
h^\beta\overline{D}_{\alpha\beta} f)\,.$$
Another important property of the (contraction) of the Riemann tensor is
$$\overline{g}^{\mu\sigma}R^\lambda_{(quad)\,\nu\lambda\sigma}= f h^\mu h^\sigma
R^\lambda_{(lin)\,\nu\lambda\sigma}\,.$$
As it can easily be seen $R^{\mu\nu}_{\ \ \rho\sigma}\equiv g^{\nu\lambda}R^\mu_{\ \rho\lambda\sigma}$ is also quadratic (and not cubic)
in $f$. More precisely we shall
write
$$R^{\mu\nu}_{\ \ \rho\sigma}=\overline{R}^{\mu\nu}_{\ \ \rho\sigma}
+R^{\mu\nu}_{(lin)\,\rho\sigma}+R^{\mu\nu}_{(quad)\rho\sigma}$$
with
$$R^{\mu\nu}_{(lin)\,\rho\sigma}=-fh^\nu h^\alpha\overline{R}^\mu_{\
\alpha\rho\sigma}
+R^{\mu\nu}_{(L)\,\rho\sigma}\quad\hbox{where}\quad R^{\mu\nu}_{(L)\rho\sigma}=\bar g^{\nu\alpha}(
\overline{D}_\rho{\Delta}^\mu_{\alpha\sigma}-\overline{D}_\sigma
{\Delta}^\mu_{\alpha\rho})\eqno({\rm A}.2)
$$
 and
$$R^{\mu\nu}_{(quad)\rho\sigma}=\overline{g}^{\nu\alpha}
R^\mu_{(quad)\,\alpha\rho\sigma}-fh^\nu h^\alpha R^\mu_{(lin)\,\alpha\rho\sigma}\,.$$
$R^{\mu\nu}_{(lin)\,\rho\sigma}$  and $R^{\mu\nu}_{(quad)\rho\sigma}$ are
antisymmetric in their lower and upper two indices.

\medskip
 One then concludes that
 the Ricci tensor (with indices up-down~: $R^\mu_\nu\equiv g^{\mu\rho}R^\lambda_{\ \rho\lambda\nu}$) is linear in $f$ [3] and reads
$$R^\mu_\nu=\overline{R}^\mu_\nu
+R^\mu_{(lin)\,\nu}\quad\hbox{with}\quad
R^\mu_{(lin)\,\nu}=-f h^\mu
h^\sigma\overline{R}_{\nu\sigma}+R^\mu_{(L)\nu}\quad\hbox{where}\quad
R^\mu_{(L)\nu}=\overline{g}^{\mu\sigma}\overline{D}_\rho
\Delta^\rho_{\nu\sigma}\,.\eqno({\rm A}.3)$$
Finally the scalar curvature is also linear in $f$ and reads
$$R=\overline{R}+R_{(lin)}\quad\hbox{with}\quad
R_{(lin)}=-f h^\alpha
h^\beta\overline{R}_{\alpha\beta}+R_{(L)}\quad\hbox{where}\quad R_{(L)}=
\overline{D}_\rho[
h^\rho\overline{D}_\mu(fh^\mu)]\,.\eqno({\rm A}.4)$$

 On can also note for further reference that
 $$R_{(L)}={1\over\sqrt{-\overline{g}}}\partial_\rho\left[h^\rho\partial_\mu\left(\sqrt{-\overline{g}}f\,h^\mu\right)\right]$$
 and that
 $$h_\alpha R^\alpha_{(L)\mu}={1\over2}fh_\alpha h^\beta\left(2h^\gamma\overline{R}^\alpha_{\ \beta\gamma\mu}+\overline{R}^\alpha_\beta\,h_\mu\right)+{h^\mu\over2}\left\{h^\alpha\partial_\alpha\left[{1\over\sqrt{-\overline{g}}}\partial_\beta\left(f\,\sqrt{-\overline{g}}h^\beta\right)\right]-f\,\overline{D}_\alpha h_\beta\overline{D}^\alpha h^\beta\right\}\,.$$

\end